\documentclass[twocolumn]{article}
\usepackage{imr}
\usepackage{amsmath, amssymb, amstext, amsthm}
\usepackage{ifpdf}

\ifpdf
\usepackage[pdftex]{graphicx}
\DeclareGraphicsExtensions{.pdf}
\else
\usepackage[dvips]{graphicx}
\DeclareGraphicsExtensions{.eps}
\fi
\def\fig#1{fig/#1}



\newif\iffig
\figtrue

\def\Real{\ensuremath{\mathbb{R}}}
\def\minW{\ensuremath{w_{\text{min}}}}
\def\S{\ensuremath{\sigma}}
\def\minS{\ensuremath{\S_{\text{min}}}}
\def\maxS{\ensuremath{\S_{\text{max}}}}
\def\minT{\ensuremath{T_{\text{min}}}}

\def\supT{\ensuremath{T_{\text{sup}}}}
\def\progS{\ensuremath{\S_{\text{prog}}}}
\def\e{\ensuremath{\varepsilon}}
\def\sp{\ensuremath{M}}
\def\diam{\text{diam}}
\def\p{\ensuremath{p}}
\def\q{\ensuremath{q}}

\def\fp{\ensuremath{P}}
\def\fq{\ensuremath{Q}}
\def\fr{\ensuremath{R}}
\def\a{\ensuremath{a}}
\def\b{\ensuremath{b}}

\def\fa{\ensuremath{A}}
\def\fb{\ensuremath{B}}

\DeclareMathOperator{\grad}{\ensuremath{\nabla}}
\def\abs#1{\ensuremath{\mathopen| #1 \mathclose|}}
\def\norm#1{\ensuremath{\mathopen\| #1 \mathclose\|}}
\def\dt{\ensuremath{\delta t}}

\def\rest#1#2{\ensuremath{\left. #1 \right|_{#2}}}
\def\next{\text{next}}
\def\half{\ensuremath{\frac{1}{2}}}

\def\ceil#1{\ensuremath{\left\lceil{#1}\right\rceil}}

\def\etal{\textsl{et al.}}

\def\aposteriori{\textsl{a posteriori}}

\def\Ungor{\"Ung\"or}

\newtheorem{lemma}{Lemma}
\newtheorem{theorem}[lemma]{Theorem}

\newtheorem{definition}{Definition}

\makeatletter
\def\begin@lgo{\begin{minipage}{1in}\begin{tabbing}
        \qquad\=\qquad\=\qquad\=\qquad\=\qquad\=\qquad\=\qquad\=\kill}
\def\end@lgo{\end{tabbing}\end{minipage}}

\newenvironment{algorithm}
{\begin{tabular}{l}\begin@lgo}
{\end@lgo\\\end{tabular}}

\makeatother

\begin{document}
%---

\title{\uppercase{%
    Efficient Spacetime Meshing
    with Nonlocal Cone Constraints%
}}

\author{%
    Shripad Thite%
}

\date{%
    Department of Computer Science, 
    University of Illinois at Urbana-Champaign;  
    thite@uiuc.edu%
}

%% Last updated July 14, 2004

% ############################################################
\abstract{%
  Spacetime Discontinuous Galerkin (DG) methods are used to solve
  hyperbolic PDEs describing wavelike physical phenomena.  When the
  PDEs are nonlinear, the speed of propagation of the phenomena,
  called the \emph{wavespeed}, at any point in the spacetime domain is
  computed as part of the solution.  We give an advancing front
  algorithm to construct a simplicial mesh of the spacetime domain
  suitable for DG solutions.  Given a simplicial mesh of a bounded
  linear or planar space domain~$\sp$, we incrementally construct a
  mesh of the spacetime domain $\sp \times [0,\infty)$ such that the
  solution can be computed in constant time per element.  We add a
  \emph{patch} of spacetime elements to the mesh at every step.  The
  boundary of every patch is \emph{causal} which means that the
  elements in the patch can be solved immediately and that the patches
  in the mesh are partially ordered by dependence.  The elements in a
  single patch are coupled because they share implicit faces; however,
  the number of elements in each patch is bounded.  The main
  contribution of this paper is sufficient constraints on the progress
  in time made by the algorithm at each step which guarantee that a
  new patch with causal boundary can be added to the mesh at every
  step even when the wavespeed is increasing discontinuously.  Our
  algorithm adapts to the local gradation of the space mesh as well as
  the wavespeed that most constrains progress at each step.  Previous
  algorithms have been restricted at each step by the maximum
  wavespeed throughout the entire spacetime domain.%
}
% ############################################################


\maketitle

\thispagestyle{empty}

% #################################################################
% #################################################################
% #################################################################

\section{Introduction}
\label{sec:intro}

Simulation problems in mechanics consider the behavior of an object or
region of space over time.  Scientists and engineers use conservation
laws and hyperbolic partial differential equations (PDEs) to model
transient, wavelike phenomena propagating over time through the domain
of interest.  Example applications are numerous, including, for
instance, the equations of elastodynamics in seismic analysis and the
Euler equations for compressible gas dynamics.  Closed-form solutions
are typically unavailable for these problems, so analysts usually
resort to numerical approximations.

Finite element methods (FEM) are popular options for solving this
class of problems.  In the standard \emph{semi-discrete} approach, a
finite element mesh discretizes space to generate a system of ordinary
differential equations in time that is then solved by a time-marching
integration scheme.  Most semi-discrete methods impose a uniform time
step size over the entire spatial domain, i.e., the time step does not
adapt to the local gradation of the space mesh.  Therefore, the
resulting spacetime mesh consists of many more elements than required
by physical causality.  Hence, algorithms that use a nonuniform time
step size can substantially improve computational efficiency.

Spacetime discontinuous Galerkin (DG) methods have been proposed by
Richter \cite{Richter94}, Lowrie \etal~\cite{Lowrie98}, and Yin
\etal~\cite{YinASHT00} for solving systems of nonlinear hyperbolic
partial differential equations.  Like traditional finite element
methods, spacetime DG methods use basis polynomials to approximate the
solution within each element; however, unlike traditional FEM methods,
these basis polynomials have local support restricted to each element
and the basis polynomials of adjacent elements do not have to agree on
their common intersection.  This approach eliminates artificial
coupling between adjacent elements when the mesh satisfies certain
causality constraints.  (For further background on general
discontinuous Galerkin methods, we refer the reader to Cockburn,
Karniadakis, and Shu~\cite{CockburnKS00}.)

\Ungor{} and Sheffer~\cite{ungor00tentpitcher} and Erickson
\etal~\cite{erickson02building} developed the first algorithm, called `TentPitcher', to build
graded spacetime meshes over arbitrary simplicially meshed spatial
domains, suitable for spacetime DG solutions.  Unlike most traditional
approaches, the TentPitcher algorithm does not impose a fixed global
time step on the mesh, or even a local time step on small regions of
the mesh.  Rather, it produces a fully unstructured simplicial
spacetime mesh, where the duration of each spacetime element depends
on the local feature size and quality of the underlying space mesh.

\begin{figure}\centering\sf
\iffig\includegraphics[height=1in]{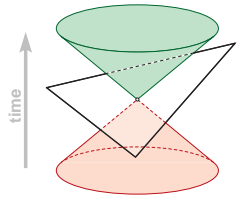}\fi
\caption{A causal face separates the cones of influence and dependence
at every point on the face.}
\label{fig:causalface}
\end{figure}

Efficient spacetime meshing relies on the notion of the domain of
influence and the domain of dependence of an event.  Imagine dropping
a pebble into a pond---circular waves propagate outwards from the
point of impact.  The frontier of expanding waves sweeps out a cone in
spacetime called the domain of influence of the event.  The radius of
the domain of influence at any time is the radius of the circular disc
consisting of all points on the surface where the initial wave has
arrived.  The domains of influence and dependence can be approximated
by right circular cones with common apex~$\fp$
(Figure~\ref{fig:causalface}).  The symmetric double cone representing
the domains of influence and dependence at points $\fp$ in spacetime
can be described by a scalar field $\omega$ where $\omega(\fp) =
\partial r / \partial t$, the \emph{wavespeed} at $\fp$, specifies how
quickly the radius $r$ of domains of influence and dependence of $\fp$
grows as a function of time.  Smaller values of $\omega(\fp)$, i.e.,
steeper cones, correspond to slower wavespeeds.  The wavespeed
$\omega(\fp)$ at a point in spacetime is, in general, part of the
solution of the PDE at that point.  The \emph{slope} of the cones of
influence and dependence of $\fp$, denoted by $\S(\fp)$, is the
reciprocal of the wavespeed---\emph{larger slopes mean steeper cones
  and therefore slower wavespeeds, and smaller slopes mean shallower
  cones and faster wavespeeds}.

\begin{figure}\centering\footnotesize\sf
\iffig\includegraphics[width=.4\textwidth]{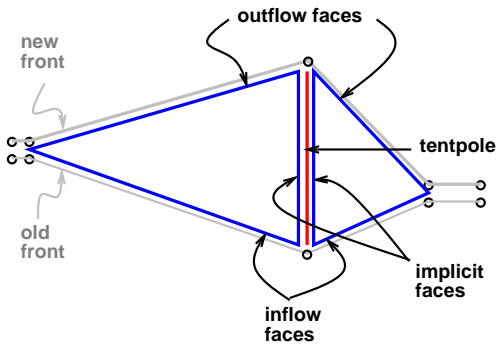}\fi
\caption{A vertical cross-section
  of a patch of tetrahedra; the inflow and outflow faces are causal.}
\label{fig:patch}
\end{figure}

Given a simplicial mesh of some bounded domain $\sp \subset \Real^d$,
the Tent Pitcher algorithm incrementally constructs a simplicial mesh
of the spacetime domain using an advancing front method.  The
spacetime domain is the subset $\sp \times [0,\infty) \subset
\Real^{d+1}$, a subset of Euclidean space one dimension higher.  The
algorithm progresses by adding simplices to the evolving mesh in small
patches by moving a vertex of the front forward in time.  The inflow
and outflow boundaries of each patch (Figure~\ref{fig:patch}) are
\emph{causal} by construction, i.e., each boundary facet $F$ separates
the cone of influence from the cone of dependence of any point on $F$
(Figure~\ref{fig:causalface}).  Equivalently, for every point $\fp$ on
$F$ we have $\norm{\grad F} \le 1/\omega(\fp) = \S(\fp)$.  If the
outflow boundaries of a patch are causal, every point in the patch
depends only on other points in the patch or points of inflow elements
adjacent to the inflow boundaries of the patch.  Therefore, the
solution within the patch can be computed as soon as the patch is
created, given only the inflow data from adjacent inflow elements.
The elements within a patch are causally dependent on each other and
must be solved as a coupled system.  Provided the space mesh has
constant degree, each patch contains only a constant number of
elements and can therefore be solved in constant time.  Therefore, the
computation time required to compute the numerical solution is
linear in the number of spacetime elements.  Patches with no causal
relationship can be solved independently.  To minimize undesirable
numerical dissipation and the number of patches, we would like the
boundary facets of each patch to be as close as possible to the
causality constraint without violating it.

The causality constraint limits the progress in time at each step,
i.e., the height of each tentpole is constrained.  For spatial domains
of dimension $d \ge 2$, it is not trivial to guarantee that the
advancing front algorithm can always make progress.  We require that
for any target time value $T$ the algorithm will compute a mesh of the
spacetime volume $\sp \times [0,T]$ and the solution everywhere in
this volume in finitely many steps.  The target time $T$ is not known
\textit{a priori} because it depends on the evolving physics.
The original Tent Pitcher algorithm proposed by \Ungor{} and
Sheffer~\cite{ungor00tentpitcher} applied to one- and two-dimensional
space domains.  The algorithm could guarantee progress only if the
input triangulation contained only angles less than 90 degrees and if
the wavespeed did not increase or increased smoothly.  Erickson
\etal~\cite{erickson02building} extended Tent Pitcher to arbitrary
spatial domains in any dimensions by imposing additional constraints,
called \emph{progress constraints}.  The progress constraint applied
to a single simplex on the front limits the amount of progress in time
when some vertex of the simplex is pitched.  The progress constraint
is a function of the shape of the simplex.  The geometric constraints
that limit the height of each tentpole are called \emph{cone
  constraints}.

All the results so far have applied to the case where the wavespeed at
a given point is either constant, decreasing, or increasing smoothly
as a Lipschitz function.  (See Alper \Ungor{}'s PhD
thesis~\cite{ungor02phd} for the details.)  When the wavespeed
changes, the previous algorithms take the fastest that the wavespeed
can ever be and use that as a conservative upper bound on the
wavespeed at any time.  One would like an algorithm that adapts to
increasing wavespeeds so that fewer spacetime elements, and therefore
less computation time, are required to mesh a given volume.

In this paper, we give an advancing front algorithm to construct a
spacetime mesh over an arbitrary linear or planar space mesh ($d \le
2$).  Our algorithm extends TentPitcher to the case when the wavespeed
can be an arbitrary scalar field over the spacetime domain.  In
particular, our algorithm guarantees finite positive progress at each
step even when the wavespeed at a given point increases
discontinuously and unpredictably over time.

The main contributions of this paper are twofold.  We give a novel
characterization of fronts that are always guaranteed to progress,
which we call \emph{progressive fronts}, and give a lower bound on the
progress guarantee at each step which depends only on the local size
of the mesh and the wavespeed that most constrains the duration of the
current patch.  The minimum progress guarantee at any step is a
positive quantity bounded away from zero, so the front is guaranteed
to progress past any target time in a finite number of steps.  The
second contribution of this paper is to give geometric constraints on
the front at any step that guarantee that the front can progress in
the next step and so on inductively at every step.  The geometric
constraints are simple to express and to compute.  Intuitively, the
geometric constraints that apply at any given iteration of the
algorithm are predicted by looking ahead at the next iteration of the
algorithm.  We also give an efficient algorithm to maximize the
progress at every step subject to these constraints.  The novelty of
our characterization of progressive fronts and of our algorithm is
that we resolve the following \emph{conundrum}.  The progress of the
front at each step $i$ is limited by the progress constraint that must
be satisfied by the next front at step $i+1$.  However, we do not know
what is the next front unless we know how much progress is possible at
step $i$.

The paper by Erickson \etal{}~\cite{erickson02building} contains an
error in the statement of the causality constraint when obtuse
triangles are involved; therefore, their proof of correctness is
incomplete because it omits the obtuse angle case.  While their proof
can be fixed, we prefer our new algorithm, which is provably correct
even when the wavespeed is constant or does not increase.  Our new
progress constraints are potentially weaker than those of Erickson
\etal{}~\cite{erickson02building}.

Our algorithm is the first algorithm to build spacetime meshes over
arbitrary planar triangulated spatial domains suitable for solving
nonlinear hyperbolic PDEs, where the wavespeed at any point in
spacetime depends on the solution and cannot be computed in advance.
Moreover, the solution can change discontinuously, for instance when a
\emph{shock} propagates through the domain.

% ...........................................................
% ...........................................................

% ...........................................................
% ...........................................................

The input to our advancing front algorithm is a simplicially meshed
bounded domain $\sp \subset \Real^d$ where $d \le 2$ and the initial
conditions of a nonlinear hyperbolic PDE.  The space mesh describes
the situation at time equal to zero, specifically, the slope at every
point in $\sp$ at time zero.  We allow more general initial conditions
but we will postpone a description of those conditions until later
sections.  Our meshing algorithm is an advancing front procedure which
alternately constructs a new patch of elements and invokes a spacetime
DG finite element method to compute the solution within that patch.
At every iteration, the \emph{front} is the graph of a continuous
piecewise linear time function $t: \sp \to \Real$.  The front $t$ is
linear within every simplex of $\sp$ and $\norm{\grad t(\p)} \le
\S(\p)$ for every point $\p \in \sp$.  The front is a terrain whose
facets correspond to simplices in the underlying space mesh.  Each
facet of the front coincides with the outflow face of a patch in the
past and the inflow face of a patch in the future.  We say that a
front is \emph{causal} if every simplex of the front is causal.  To
advance the front $t$, the algorithm chooses an arbitrary vertex $\fp
= (\p, t(\p))$ from the front and lifts it to a new point $\fp' = (\p,
t'(\p))$ where $t'(\p) > t(\p)$ and for every other vertex $\q$ we
have $t'(\q) = t(\q)$.  The spacetime volume between the new front
$t'$ and the old front $t$ is called a \emph{tent}.  The tent is
meshed with simplices sharing the edge $(\fp, \fp')$ called the
\emph{tentpole}.  The \emph{height} of the tentpole is the duration
$t'(\p) - t(\p)$.  Consider a planar space mesh $\sp$.  For each
triangle $pqr$ incident on $p$, the tetrahedron $\fp'\fp\fq\fr$
belongs to the patch.  The outflow face $\fp'\fq\fr$ and the inflow
face $\fp\fq\fr$ are causal boundaries.  The triangles $\fp'\fp\fq$
and $\fp'\fp\fr$ are implicit faces.  Since the implicit faces are
vertical they are not causal boundaries and so elements within the
patch are coupled.  The elements below the front $t$ whose outflow
faces intersect any of the inflow faces of the new patch are inflow
elements.  We pass the newly constructed patch along with all its
inflow elements to a DG solver.  The DG solver returns as part of the
solution the slope at every point on every outflow face of the patch.
The new front $t'$ and the output of the DG solver are the input to
the next iteration of the algorithm.

\emph{Since we are interested in causal fronts only, henceforth it is
  implicit that every front considered is causal.}

We assume that the slope at any point~$\fp$ is bounded by the minimum
and the maximum slopes anywhere in the cone of dependence of~$\fp$.
Hence, given a front~$t$ and a point~$\fp$ in the future, the slope
at~$\fp$ is no smaller than the slope at~$\fq$ for every point~$\fq$
on the front~$t$ such that~$\fp$ is in the cone of influence of~$\fq$.

It can be computationally very expensive to determine the shallowest
cone of influence that contains a given point~$\fp$.  In particular,
the shallowest cone of influence containing~$\fp$ may correspond to a
\emph{nonlocal} point~$\fq$, one arbitrarily distant from~$\fp$.  To
compute this nonlocal cone constraint efficiently, we use a standard
hierarchical decomposition, called a \emph{bounding cone hierarchy},
of the space domain.  The elements in the hierarchy correspond to
subsets of the space domain.  For each element of the hierarchy, we
compute the minimum slope within the corresponding subset of the space
domain.  The smallest element in the hierarchy is a single simplex.
In order to determine the strictest cone constraint that applies
locally, we traverse the hierarchy until we determine the simplex with
minimum slope whose cone of influence contains~$\fp$.  In practice, we
expect that our algorithm has to examine only a small subset of the
hierarchy.  In the worst case, the algorithm has to examine every
simplex of the front but in that case the algorithm will be at most a
constant factor slower than one that does not use a bounding cone
hierarchy.  When a patch is solved, the bounding cones are updated
with the new slopes by traversing a path from a leaf to the root of
the hierarchy.  This hierarchical approximation technique has been
applied very successfully to numerous simulation problems, such as the
Barnes-Hut divide-and-conquer method~\cite{barnes-hut86nbody} for
$N$-body simulations, as well as to collision detection in computer
graphics and robot motion planning~\cite{lin96collision} and for
indexing multi-dimensional data in geographic information
systems~\cite{guttman84rtrees}.

\iffalse
In Section~\ref{sec:1d}, we describe the problem and our solution for
the case of one-dimensional space domains.  Several aspects of the
complexity of the problem are evident even in the 1D$\times$Time case.
In Section~\ref{sec:2d}, we describe our algorithm for planar space
domains.  We describe sufficient conditions such that the front at
every step is guaranteed to make progress.  Finally, we conclude by
comparing our results to previous work on the uniform wavespeed case.
\fi

% ================================================================
% ================================================================
% ================================================================
\subsection{Notation}

We use lowercase letters like $p$, $q$, $r$ to denote points in space
and uppercase letters like $\fp$, $\fq$, $\fr$ to denote points in
spacetime.  A front $t$ is a piecewise linear function $t : \sp \to
\Real$.  For a simplex (of any dimensions) $\tau$ of $\sp$, let
$\rest{t}{\tau}$ denote the time function $t$ restricted to $\tau$ and
extended to the affine hull of $\tau$; in other words,
$\rest{t}{\tau}$ is a linear function that coincides with $t$ for
every point of $\tau$.  Let $t_i : \sp \to \Real$ denote the front after
the $i$th step of the algorithm; $t_0$ is the initial front.  For
every~$i$, the front~$t_i$ is a terrain whose facets are the simplices
of~$\sp$.  In other words, $t_i$ is a piecewise linear function such
that for every simplex~$\tau$ of~$M$, the functions~$t_i$
and~$\rest{t_i}{\tau}$ coincide at the vertices of~$\tau$.

For a time function~$t:\sp \to \Real$ we denote the gradient of~$t$
by~$\grad t$.  A local minimum of the front~$t$ is a vertex~$p$ such
that~$t(p) \le t(q)$ for every vertex $q$ that is a neighbor of $p$.
When the current front $t$ is clear from the context, for every point
$p \in \sp$ we use $\fp$ to denote the corresponding point on the
front, i.e., $\fp=(p,t(p))$.

For a point~$\fp$ in spacetime, we use~$\S(\fp)$ to denote the
reciprocal of the wavespeed at $\fp$.  Let $\minS$ denote $\min_{\fp
  \in \sp \times [0,\infty)} \{\S(\fp)\}$ and $\maxS$ denote
$\max_{\fp \in \sp \times [0,\infty)} \{\S(\fp)\}$.  We assume that $0
< \minS \le \maxS < \infty$.  For a simplex~$\tau$ in spacetime, we
use~$\S(\tau)$ to denote the minimum of~$\S(\fp)$ over all
points~$\fp$ in~$\tau$.

We say that a front~$t'$ is obtained by advancing a vertex~$p$ of~$M$
by~$\dt \ge 0$ if~$t'(p) = t(p) + \dt$ and for every other vertex~$q
\ne p$ we have~$t'(q) = t(q)$.  For any front $t$, vertex $\p$, and
real $\dt \ge 0$, let $t'=\next(t,\p,\dt)$ denote the front obtained
from $t$ by advancing $\p$ by $\dt$.

% ================================================================
% ================================================================
% ================================================================
\subsection{Problem statement}

The input to our problem is the initial front~$t_0$ and the initial
conditions of the PDE.  We want an advancing front algorithm such that
for every~$T \in \Real^{\ge 0}$ there exists a finite integer $k \ge 0$
such that the front $t_k$ after the $k$th iteration of the algorithm
satisfies $t_k \ge T$.

We say that a front~$t$ is \emph{valid} if there exists a positive
real $\delta$ bounded away from zero such that for every~$T \in
\Real^{\ge 0}$ there exists a sequence of fronts $t$, $t_1$, $t_2$,
$\ldots$, $t_k$ where $t_k \ge T$, each front in the sequence obtained
from the previous front by advancing some vertex by $\delta$.  What
makes the definition of a valid front nontrivial is the requirement
that all fronts be causal.  The main difficulty in characterizing
valid fronts arises when the wavespeed at a given point in the space
domain increases discontinuously and unpredictably over time.

\noindent\textbf{Our solution}
We define \emph{progressive} fronts and prove that if a front is
progressive then it is valid.  We give an algorithm that given any
progressive front $t_i$ constructs a next front~$t_{i+1}$ such that
$t_{i+1}$ is progressive.  The volume between $t_i$ and $t_{i+1}$ is
partitioned into simplices.  The next front $t_{i+1}$ is
obtained by lifting a local minimum of $t_i$ by a positive amount
bounded away from zero.  The algorithm can easily be parallelized to
solve several patches asynchronously by lifting any independent set of
vertices in parallel.  Whenever the algorithm chooses to lift a local
minimum, it is guaranteed to be able to lift it by at least $\minT >
0$ which is a function of the input and bounded away from zero.

%% End introduction.

% #################################################################
% #################################################################
% #################################################################

\section{One-dimensional space domains}
\label{sec:1d}

We begin by describing our algorithm to construct spacetime meshes
over one-dimensional space domains.  Even this simple case captures
all but one aspect of the complexity of guaranteeing causality when
wavespeeds are changing.

The space domain $\sp$ is a closed interval of the real line.  The
input space mesh is a subdivision of this interval into segments.  Let
$V(\sp)$ denote the set of vertices of the space mesh $\sp$.  The
initial front $t_0$ corresponds to $t_0(p) = 0$ for every vertex $p$ of
the space mesh, but more generally, any (causal) front can be the
initial front.  Let $\minW$ denote the minimum length of any segment
in the space mesh.  Let $\minS$ denote the minimum slope $\S(\fp)$
over every point $\fp$ in the spacetime domain $\sp \times
[0,\infty)$.  Let $\minT$ denote $\minS \minW$.

In iteration $i+1$ of our advancing front algorithm ($i \ge 0$), we
advance a single vertex~$p$, where $p$ is a local minimum of the
current front~$t_i$, to get the new front~$t_{i+1}$, i.e., $t_{i+1} =
\next(t,p,\dt)$.  More generally, we can advance any vertex or an
independent set of vertices, not necessarily local minima, forward in
time.  The value of $t_{i+1}(p)$ is bounded from above by the
requirement that $t_{i+1}$ be causal.

Let~$\fa\fb$ be an arbitrary segment of the front~$t_{i+1}$. Without
loss of generality, assume~$t_{i+1}(\a) \le t_{i+1}(\b)$.  Then,
$\fa\fb$ is causal if and only if the gradient of the time function
$t_{i+1}$ restricted to $ab$ is at most the slope $\S(\fa\fb)$, i.e.,
if and only if
  \begin{equation}\boxed{%
    \norm{\grad \rest{t_{i+1}}{ab}}
  =
    \frac{t_{i+1}(b) - t_{i+1}(a)}{\abs{ab}}
  \le
    \S(\fa\fb).
  \label{eqn:1d:causalityconstraint}
  }\end{equation}

% ................................................................
\begin{theorem}
  Let $t_i$ be a front and let $p$ be an
  arbitrary local minimum of $t_i$.  Then, for every $\dt \in [0,\minT]$
  the front $t_{i+1} = \next(t_i,p,\dt)$ is causal.
\label{thm:1d:causal}
\end{theorem}
\begin{proof}
  Only the segments of the front incident on $\fp$ advance along
  with~$p$.  Consider an arbitrary segment $pq$ incident on $p$.
  Let~$t$ and~$t'$ denote~$\rest{t_i}{pq}$ and~$\rest{t_{i+1}}{pq}$
  respectively.  We have $t(p) + \dt \le t(p) + \minS \minW \le t(q) +
  \abs{pq} \S(\fp'\fq)$ because $p$ is a local minimum, $\minW \le
  \abs{pq}$, and $\minS \le \S(\fp'\fq)$.  Therefore, the segment
  $\fp'\fq$ is causal.  Since this is true of an arbitrary segment on
  the front $t'$, we have proved that that the front $t_{i+1} =
  \next(t_i,p,\dt)$ is causal.
\end{proof}
% ................................................................

% ................................................................
\begin{theorem}
  For any $i \ge 0$, if the front $t_i$ is causal then $t_i$ is
  valid.
\label{thm:1d:causalisvalid}
\end{theorem}
\begin{proof}
  Consider step $i+1$ of the algorithm.  By
  Theorem~\ref{thm:1d:causal} the front $t_{i+1}$ such that
  $t_{i+1}(p) \in [0,\minT]$ is causal.  Therefore, we have shown that
  if $t_i$ is causal then there is a front $t_{i+1} =
  \next(t_i,p,\minT)$ such that $t_{i+1}$ is causal.  Note that
  $\sum_{p \in V(\sp)} t_{i+1}(p) = \minT + \sum_{p \in V(\sp)}
  t_i(p)$.  By induction on $i$, and because $\maxS$ is finite and
  $\sp$ is bounded, there exists a finite $k \ge i$ such that the
  front $t_k$ satisfies
  \[
    \sum_{p \in V(\sp)} t_k(p)
  \ge
   \diam(\sp) \maxS T
  \]
  for any real $T$.  Since $t_k$ is causal
  \[
    \left( \max_{p \in V(\sp)} t_k(p) \right)
  \le
    \diam(\sp) \maxS \left( \min_{p \in V(\sp)} t_k(p) \right).
  \]
  Therefore, $\min_{p \in V(\sp)} t_k(p) \ge T$ and so $t_i$ is valid.
\end{proof}
% ................................................................

% ================================================================
% ================================================================
% ================================================================
\subsection{Being greedy at every step}

\begin{figure}
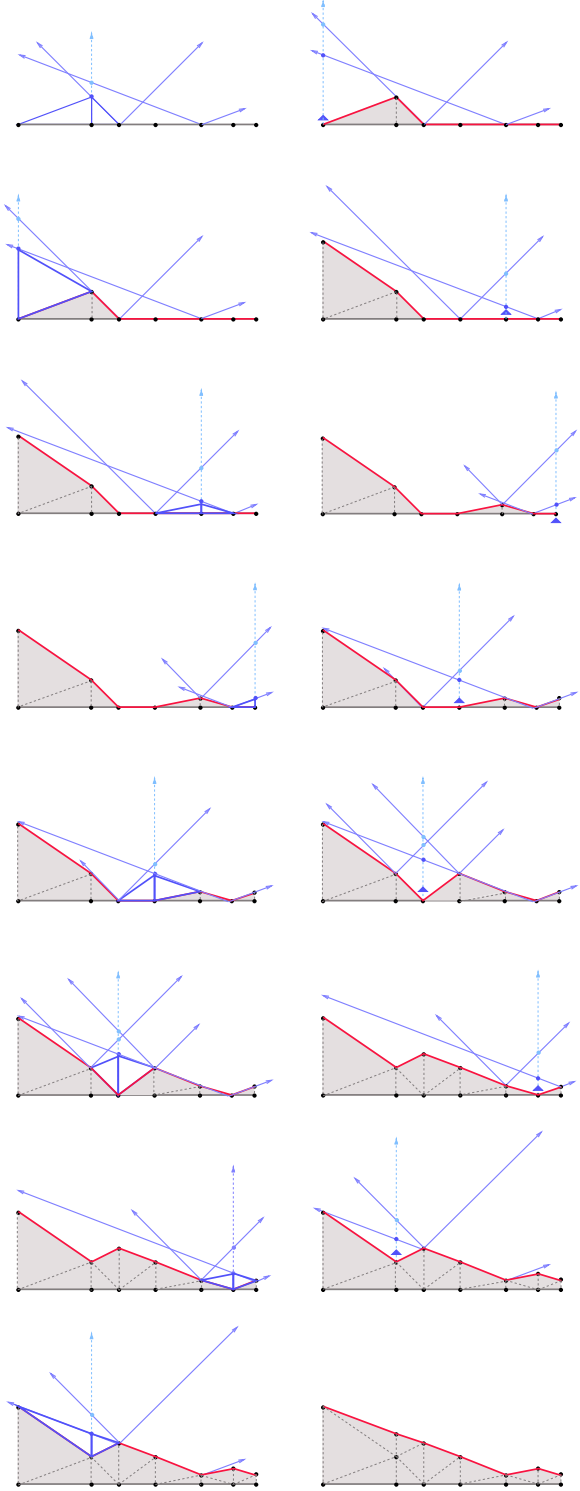
\centering\sf
\begin{tabular}{cc}
\iffig\includegraphics[width=0.22\textwidth]{\fig{mesh-1}}\fi&
\iffig\includegraphics[width=0.22\textwidth]{\fig{mesh-2}}\fi\\
\iffig\includegraphics[width=0.22\textwidth]{\fig{mesh-3}}\fi&
\iffig\includegraphics[width=0.22\textwidth]{\fig{mesh-4}}\fi\\
\iffig\includegraphics[width=0.22\textwidth]{\fig{mesh-5}}\fi&
\iffig\includegraphics[width=0.22\textwidth]{\fig{mesh-6}}\fi\\
\iffig\includegraphics[width=0.22\textwidth]{\fig{mesh-7}}\fi&
\iffig\includegraphics[width=0.22\textwidth]{\fig{mesh-8}}\fi\\
\iffig\includegraphics[width=0.22\textwidth]{\fig{mesh-9}}\fi&
\iffig\includegraphics[width=0.22\textwidth]{\fig{mesh-10}}\fi\\
\iffig\includegraphics[width=0.22\textwidth]{\fig{mesh-11}}\fi&
\iffig\includegraphics[width=0.22\textwidth]{\fig{mesh-12}}\fi\\
\iffig\includegraphics[width=0.22\textwidth]{\fig{mesh-13}}\fi&
\iffig\includegraphics[width=0.22\textwidth]{\fig{mesh-14}}\fi\\
\iffig\includegraphics[width=0.22\textwidth]{\fig{mesh-15}}\fi&
\iffig\includegraphics[width=0.22\textwidth]{\fig{mesh-16}}\fi
\end{tabular}
\caption{Top to bottom: a sequence of tent pitching steps in
  1D$\times$Time.  Maximizing the height of each tentpole while
  staying below every cone of influence can require examining remote
  cones arbitrarily far away.}
\label{fig:1d:tents}
\end{figure}

We would like to maximize the progress at each step in a greedy
fashion, i.e., given a front $t_i$ we would like to maximize
$t_{i+1}(p)$ where $t_{i+1} = \next(t_i,p,\dt)$ subject to the
constraint that $t_{i+1}$ is causal.  By
Theorem~\ref{thm:1d:causalisvalid}, we can have $t_{i+1}(p) \ge t_i(p)
+ \minT$.  However, it may be possible to make further progress by
setting $t_{i+1}(p)$ higher, especially if each segment $\fp\fq$ incident on
$p$ each satisfies progress constraint [$\S_{\text{prev}}$] for some
$\S_{\text{prev}} < \S(\fp'\fq)$ at the end of the previous iteration.

For a fixed segment $pq$ incident on $p$ let $\supT^{i+1}$ denote
$\sup \, \{ T : \fp'\fq$ is causal where $\fp'=(p,T) \}$.  To
maximizing the progress at step $i+1$, we would like to compute
$\supT^{i+1}$.  The segment $\fp'\fq$ is causal if and only if the
slope of $\fp'\fq$ is less than or equal to the slope of the cone of
influence from every point on the front that intersects $\fp'\fq$.  A
cone of influence intersects $\fp'\fq$ if and only if the cone
intersects the tentpole $\fp\fp'$.  In general, a cone of influence
from arbitrarily far away can intersect the tentpole at $p$. See
Figure~\ref{fig:1d:tents}.  This is not the case when the wavespeed
everywhere is the same.  Therefore, in general, $\supT^{i+1}$ could be
determined by a cone of influence of a point arbitrarily distant from
$p$.

Partition the front into two subsets of points: (i)~points in the star
of $\fp$ (``local'' points), and (ii)~points everywhere else on the
front (``remote'' points).  Corresponding to each subset we have two
disjoint subsets of cones of influence---$\mathcal{C}_{\text{local}}$
and $\mathcal{C}_{\text{remote}}$ respectively.  Each subset of cones
limits the new time value of $p$ and so the final time value is the
smaller of the two values for each of $\mathcal{C}_{\text{local}}$ and
$\mathcal{C}_{\text{remote}}$ taken separately.

Consider the subset $\mathcal{C}_{\text{local}}$.  Let
$\S_{\text{local}}$ denote the smallest slope among all cones of
influence in $\mathcal{C}_{\text{local}}$ The segment $\fp'\fq$ is
causal only if its slope is less than or equal to $\S_{\text{local}}$.
Let $T_{\text{local}}$ be the maximum time value of $\fp'$ for which
the slope of $\fp'\fq$ is less than or equal to $\S_{\text{local}}$.
The maximum $T_{\text{local}}$ exists because the set of feasible
values is closed and therefore compact.  To compute $T_{\text{local}}$
we substitute $\S_{\text{local}}$ in the condition for causality of
$\fp'\fq$ (Equation~\ref{eqn:1d:causalityconstraint}).

Next consider the subset $\mathcal{C}_{\text{remote}}$.  The front
$t_i$ is strictly below every cone in $\mathcal{C}_{\text{remote}}$
because $t_i$ is causal.  The segment $\fp'\fq$ is causal only if it
is also strictly below every cone in $\mathcal{C}_{\text{remote}}$.
Given a cone $C \in \mathcal{C}_{\text{remote}}$, $C$ intersects
$\fp'\fq$ if and only if $C$ intersects the tentpole $\fp\fp'$.  Let
$T_{\text{remote}}$ denote the smallest time value $T$ for which the
tentpole $\fp\fp'$ where $\fp'=(p,T)$ intersects exactly one cone in
$\mathcal{C}_{\text{remote}}$.    The segment $\fp'\fq$ is causal only
if $T < T_{\text{remote}}$.  Note that the upper bound on $T$ imposed
by remote cones is a strict inequality.

Therefore, the progress $t_{i+1}(p) - t_i(p)$ at step $i+1$ is limited
because $\supT^{i+1} = \max \{T_{\text{local}}, T_{\text{remote}}\}$.
To maximize the progress at the current step, we choose $t_{i+1}(p)$
equal to $\supT^{i+1}$ minus the machine precision $\eta$, or $t_i(p)
+ \minT$, whichever is larger.

% ----------------------------------------------------------------
% ----------------------------------------------------------------
% ----------------------------------------------------------------
\paragraph{Computing $\mathbf{T_{\text{remote}}}$ exactly}

Computing $T_{\text{remote}}$ is equivalent to answering a ray
shooting query in the arrangement of the cones in
$\mathcal{C}_{\text{remote}}$.  We use a bounding cone hierarchy
$\mathcal{H}$ obtained from a hierarchical decomposition of the space
domain to efficiently answer the ray shooting query.  The hierarchical
decomposition of the space domain induces a corresponding hierarchical
decomposition of every front.  For each element of this hierarchy, we
store a right circular cone that bounds the cone of influence of every
point of the corresponding subset of the front.  To answer the ray
shooting query, we traverse the cone hierarchy from top to bottom
starting at the root.  At every stage, we store a subset $\mathcal{C}$
of bounding cones such that every cone in
$\mathcal{C}_{\text{remote}}$ is contained in some cone in the subset
$\mathcal{C}$.  The cones in $\mathcal{C}$ are stored in a priority
queue in non-decreasing order of the time value at which the vertical
ray at $\fp$ intersects each cone.  Initially, $\mathcal{C}$ consists
solely of the cone at the root of the hierarchy.  At every stage, if
the cone in $\mathcal{C}$ that has the earliest intersection does not
come from a leaf in the hierarchy then we replace it in the priority
queue with its children.  Continuing in this fashion, we eventually
determine the single facet of the front such that the cone of
influence from some point on this facet is intersected first by the
vertical ray at $\fp$.  The time coordinate of the point of
intersection is $T_{\text{remote}}$, the answer to the ray shooting
query.

If the hierarchy is balanced its depth is $O(\log m)$ where $m$ is the
number of simplices in the space mesh.  In 1D$\times$Time, we observed
empirically that on average only a few nodes in the cone hierarchy
were examined by this algorithm to determine the most constraining
cone of influence.

% ----------------------------------------------------------------
% ----------------------------------------------------------------
% ----------------------------------------------------------------
\paragraph{Approximating $\mathbf{T_{\text{remote}}}$}

Since we know a range of values $[t_i(p) + \minT, T_{\text{local}}]$
that contains $T_{\text{remote}}$, we can approximate
$T_{\text{remote}}$ up to any desired numerical accuracy by performing
a binary search in this interval.  At every iteration, we
speculatively lift $\fp$ to the midpoint of the current search
interval.  Let $\fp''$ be the speculative top of the tentpole at
$\fp$.  We query the cones of influence in
$\mathcal{C}_{\text{remote}}$ to determine the minimum slope
$\S_{\text{remote}}$ among all cones that intersect $\fp\fp''$.  If
the maximum slope of the outflow faces incident on $\fp''$ is less
than $\S_{\text{remote}}$ then we can continue searching in the top
half of the current interval; otherwise, the binary search continues
in the bottom half of the current interval.  The search terminates
when the search interval is smaller than our desired accuracy.  A
bounding cone hierarchy helps in the same manner as before to
determine the minimum slope among all cones in
$\mathcal{C}_{\text{remote}}$ that intersect $\fp\fp''$.

%\bigskip

% .................................................................
\begin{theorem}
  Given a simplicial mesh $\sp$ of a bounded real interval where
  $\minW$ is the minimum length of a simplex of $\sp$ and $\minS$ is
  the minimum slope anywhere in $\sp \times [0,\infty)$ our algorithm
  constructs a simplicial mesh of~$\sp \times [0,T]$ consisting of at
  most~$\ceil{\frac{2 \, \diam(\sp) \, \maxS}{\minS \minW}
    \, T}$ spacetime elements for every real $T \ge 0$.
\label{thm:1d:main}
\end{theorem}
\begin{proof}
  In Theorem~\ref{thm:1d:causal}, we have shown that the height of
  each tentpole constructed by the algorithm is at least $\minT =
  \minS \minW$.  By Theorem~\ref{thm:1d:causalisvalid}, after
  constructing at most $k \le \ceil{\frac{\diam(\sp) \, \maxS}{\minT}
    \, T}$ patches, the entire front $t_k$ is past the target time
  $T$.  Since each patch consists of at most two elements, the theorem
  follows.
\end{proof}
% ..................................................................

We have shown that every causal front in 1D$\times$Time is valid.  In
higher dimensions, additional progress constraints are necessary.

%% End algorithm for 1D x Time

% #################################################################
% #################################################################
% #################################################################

\section{Planar space domains}
\label{sec:2d}

In this section, we describe our algorithm for $d=2$, i.e., for a
triangulated planar space domain $\sp \subset \Real^2$.

For planar domains, we encounter nontrivial progress constraints that
are necessary to guarantee sufficient progress at each step, i.e., to
guarantee that the height of the tentpole constructed at every step is
positive and bounded away from zero.  In the absence of such
constraints, it was shown by \Ungor{} and
Sheffer~\cite{ungor00tentpitcher}, and by Erickson
\etal{}~\cite{erickson02building} that if the space mesh contains an
obtuse or a right triangle then Tent Pitcher will eventually construct
a front such that no further progress is possible while maintaining
causality.  Erickson \etal{}~\cite{erickson02building} derived
additional progress constraints that were sufficient to guarantee
progress, even in the presence of obtuse angles, however only by
assuming the minimum slope occurs everywhere in spacetime.  In this
section, we show how to relax these progress constraints so that they
adapt to the slope of the most constraining cone of influence at every
step.  Our progress constraint is a function of the slope encountered
locally in the next step of the algorithm, which may be substantially
less constraining than the globally minimum slope.

Fix a real parameter $\e \in \left(0,\half\right]$.  The space domain
$\sp$ is a triangulation of a bounded subset of the plane $\Real^2$.
Let $\minW$ denote the minimum width of any triangle of the space
mesh.  Let $\minS$ denote the minimum $\S(\fp)$ over every point $\fp$
in the spacetime domain $\sp \times [0,\infty)$.  Let $\minT$ denote
$\e \minS \minW$.

\begin{definition}[Progress constraint~$\S$]
  Let~$\fp\fq\fr$ be an arbitrary triangle of a front~$t$. Without
  loss of generality, assume~$t(p) \le t(q) \le t(r)$.  We say that
  the triangle~$\fp\fq\fr$ satisfies \emph{progress constraint~$\S$}
  if and only if
  \[ 
    \norm{\grad \rest{t}{qr}}
  :=
    \frac{t(r) - t(q)}{\abs{qr}}
  \le 
    (1-\e) \S \phi_p
  \]
  where $\phi_p = \max\,\{\sin \angle{prq}, \sin \angle{pqr}\}$.
  Note that $0 < \phi_p \le 1$.
\end{definition}

Suppose the lowest vertex $p$ is being advanced.  As long as $p$ is
the lowest vertex of $\triangle{pqr}$, the progress constraint limits
$\norm{\grad \rest{t}{qr}}$ but $\norm{\grad \rest{t}{qr}}$ is
unchanged by lifting $p$.  When $t(p) > t(q)$, the new lowest vertex
is $q$, so the progress constraint limits $\norm{\grad \rest{t}{rp}}$.
(We can interpret the progress constraint inductively as a
causality constraint on the $1$-dimensional facet $pr$ opposite $q$
where the relevant slope is $(1-\e) \S \phi_q$.)

\begin{definition}[Progressive]
  Let $t$ be a front and let $pqr$ be a given
  triangle.  Without loss of generality, assume~$t(p) \le t(q)
  \le t(r)$.  We say that the triangle
  $\fp\fq\fr$ is \emph{progressive} if and only if
  both of the following conditions are satisfied by
  $\fp'\fq\fr$ where $\fp' = (p, t(p) + \dt)$
  for every $\dt \in [0,\minT]$:
  \begin{enumerate}
  \item $\fp'\fq\fr$ is causal, and
  \item $\fp'\fq\fr$ satisfies progress constraint
    $\S(\fp'\fq'\fr)$ where $\fq' = (q, t(q) + \minT)$.
  \end{enumerate}
\end{definition}

We say that a front $t$ is \emph{progressive} if every triangle on the
front is progressive.  Note that every progressive triangle or front
is also causal.

% ===============================================================
% ===============================================================
% ===============================================================

\subsection{A new advancing front algorithm}

We are now ready to describe iteration $i+1$ of our advancing front
algorithm for $i \ge 0$.  Advance a single vertex~$p$ by a positive
amount, where $p$ is any local minimum of the current front~$t_i$, to
get the new front~$t_{i+1}$ such that for every triangle $pqr$
incident on $p$ the corresponding triangle on the new front $t_{i+1}$
is progressive.  In the parallel setting, advance any independent set
of local minima forward in time, each subject to the above constraint.
The value of $t_{i+1}(p)$ is constrained from above separately for
each of the simplices incident on $p$.  The final value chosen by the
algorithm must satisfy the constraints for each such triangle.
Therefore, it is sufficient to consider each triangle $pqr$ incident
on $p$ separately while deriving the causality and progress
constraints that apply while pitching $p$.

Next, we derive simple formul\ae{} for the causality and progress
constraints for a given triangle $pqr$ when $p$ is
being pitched.  Let~$t$ and~$t'$
denote~$\rest{t_i}{pqr}$
and~$\rest{t_{i+1}}{pqr}$ respectively.

Let~$\vec{n}_{qr}$ denote the unit vector normal to~$qr$ such
that~$\vec{n}_{qr} \cdot (\vec{p} - \vec{q}) > 0$.  Let $\vec{v}_{qr}$
be the unit vector parallel to $qr$ such that $\vec{v}_{qr} \cdot
(\vec{r} - \vec{q}) > 0$.  Then, $\{\vec{n}_{qr}, \vec{v}_{qr}\}$ form
a basis for the vector space $\Real^2$.  Let~$\vec{n}_{rp}$ denote the
unit vector normal to~$pr$ such that~$\vec{n}_{rp} \cdot (\vec{q} -
\vec{p}) > 0$.  Let $\vec{v}_{rp}$ be the unit vector parallel to $rp$
such that $\vec{rp} \cdot (\vec{p} - \vec{r}) > 0$.  Then,
$\{\vec{n}_{rp}, \vec{v}_{rp}\}$ form another basis for the vector
space $\Real^2$.

The gradient vector $\grad t'$ can be written as
\[
  \grad t'
=
  (\grad t' \cdot \vec{n}_{qr}) \vec{n}_{qr}
+
  \grad \rest{t'}{qr}
\]
where
\[
  \grad \rest{t'}{qr}
=
  (\grad t' \cdot \vec{v}_{qr}) \vec{v}_{qr}
\]
Lifting $p$ does not change the gradient of the time function
restricted to the opposite edge, so $\grad t' \cdot \vec{v}_{qr} =
\grad t \cdot \vec{v}_{qr}$, i.e., $\grad \rest{t'}{qr} = \grad
\rest{t}{qr}$.  Since $q$ is the lowest vertex of $qr$, we have $\grad
t' \cdot \vec{v}_{qr} = \grad t \cdot \vec{v}_{qr} \ge 0$.

Also,
\[
  \grad t'
=
  (\grad t' \cdot \vec{n}_{rp}) \vec{n}_{rp}
+
  \grad \rest{t'}{rp}
\]
where
\[
  \grad \rest{t'}{rp}
=
  (\grad t' \cdot \vec{v}_{rp}) \vec{v}_{rp}
\]
The vectors $\vec{n}_{qr}$ and $\vec{n}_{rp}$ are related by a rotation
around the origin by angle $\theta$.  Since $0 < \theta < \pi$ we have
$\cos \theta = \vec{n}_{qr} \cdot \vec{n}_{rp}$ and $\sin \theta = \sqrt{1 -
  (\vec{n}_{qr} \cdot \vec{n}_{rp})^2}$.  Hence,
\begin{align}
  \norm{\grad \rest{t'}{rp}}
&=
  \norm{\grad \rest{t}{qr}} \, \cos \theta
+
  (\grad t' \cdot \vec{n}_{qr}) \, \sin \theta \nonumber\\
&=
  \norm{\grad \rest{t}{qr}} (\vec{n}_{qr} \cdot \vec{n}_{rp}) \nonumber\\
&\quad {}+{}
  (\grad t' \cdot \vec{n}_{qr}) \sqrt{1 - (\vec{n}_{qr} \cdot \vec{n}_{rp})^2}
\label{eqn:2d:rotatebytheta}
\end{align}

% ================================================================
% ================================================================
% ================================================================
\paragraph{Deriving the causality constraint}

Let $u$ be the orthogonal projection of $p$ onto line $qr$.
Since lifting $p$ does not change the time function
restricted to $qr$, we have $\rest{t'}{qr} =
\rest{t}{qr}$.  The scalar product~$\grad t' \cdot \vec{n}_{qr}$ can be
written as
\[
  \grad t' \cdot \vec{n}_{qr}
=
  \frac{t'(p)-t(u)}{\abs{up}}
\]
Since $q$ is the lowest vertex of $qr$ and since
$\fp\fq\fr$ is progressive, we have $0 \le \grad t'
\cdot \vec{v}_{qr} = \grad t \cdot \vec{v}_{qr} \le (1-\e)
\S(\fp'\fq\fr) < \S(\fp'\fq\fr)$.
Therefore, $\norm{\grad t'} \le
\S(\fp'\fq\fr)$ if and only if
\begin{equation}\boxed{%
  \frac{t'(p) - t(u)}{\abs{up}}
\le 
  \sqrt{\S(\fp'\fq\fr)^2 - \norm{\grad \rest{t}{qr}}^2}
\label{eqn:2d:causalityconstraint}
}\end{equation}

% ================================================================
% ================================================================
% ================================================================
\paragraph{Deriving the progress constraint}

Let $\progS$ denote $\S(\fp'\fq'\fr)$ where $\fp'=(p,t'(p))$ and
$\fq'=(q,t(q)+\minT)$.  By Equation~\ref{eqn:2d:rotatebytheta}, the
triangle $\fp'\fq\fr$ satisfies the progress constraint $\norm{\grad
  \rest{t'}{rp}} \le (1-\e) \progS \phi_q$ if and only if
\[
  \grad t' \cdot \vec{n}_{qr}
\le
  \frac{(1-\e) \progS \phi_q
        - \norm{\grad \rest{t}{qr}} (\vec{n}_{qr} \cdot \vec{n}_{rp})}
       {\sqrt{1 - (\vec{n}_{qr} \cdot \vec{n}_{rp})^2}}
\]
Therefore, the progress constraint is
\begin{equation}\boxed{%
\begin{array}{rcl}
  \frac{t'(p) - t(u)}{\abs{up}}
&\le&
  \frac{1}{\sqrt{1 - (\vec{n}_{qr} \cdot \vec{n}_{rp})^2}}
  \, (1-\e) \S(\fp'\fq'\fr) \phi_q \\[2ex]
&&
  {}-{} \frac{\vec{n}_{qr} \cdot \vec{n}_{rp}}
             {\sqrt{1 - (\vec{n}_{qr} \cdot \vec{n}_{rp})^2}} 
  \, \norm{\grad \rest{t}{qr}}
\end{array}
\label{eqn:2d:progressconstraint}
}\end{equation}

% ================================================================
% ================================================================
% ================================================================

\subsection{Proof of correctness}

In this section, we prove the correctness of our algorithm, i.e., that
every front constructed by the algorithm is valid.

% ..................................................................
\begin{figure}
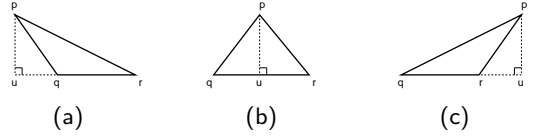
\centering\sf
\begin{tabular}{ccc}
\iffig\includegraphics[width=0.13\textwidth]{\fig{pqr-Qobtuse}}\fi
&
\iffig\includegraphics[width=0.13\textwidth]{\fig{pqr-QRacute}}\fi
&
\iffig\includegraphics[width=0.13\textwidth]{\fig{pqr-Robtuse}}\fi\\
(a) & (b) & (c)
\end{tabular}
\caption{Triangle $pqr$ where $t(p) \le t(q) \le t(r)$}
\label{fig:pqr}
\label{fig:pqr-Qobtuse}
\label{fig:pqr-QRacute}
\label{fig:pqr-Robtuse}
\end{figure}
% ..................................................................

% -----------------------------------------------------------------
% -----------------------------------------------------------------
\begin{theorem}
  If a front $t_i$ is progressive, then for any local minimum vertex
  $p$ and for every $\dt \in [0,\minT]$ the front $t_{i+1} =
  \next(t_i,p,\dt)$ is causal.
\label{thm:2d:iscausal}
\end{theorem}
\begin{proof}
Since only the triangles of the front incident on~$\fp$ advance
along with~$p$, we can restrict our attention to an arbitrary
triangle~$pqr$ incident on~$p$.  Let~$t$
and~$t'$ denote~$\rest{t_i}{pqr}$
and~$\rest{t_{i+1}}{pqr}$ respectively.
Let $u$ be the orthogonal projection of $p$ onto line $qr$.

Consider the causality constraint
(Equation~\ref{eqn:2d:causalityconstraint}).  We will consider two cases
separately: (i)~$t(u) \ge t(q)$, and (ii)~$t(u) < t(q)$.

% ..................................................................
\noindent\textbf{Case 1: $\mathbf{t(u) \ge t(q) \ge t(p)}$}
See Figure~\ref{fig:pqr-QRacute}(b)--(c).  In this case, we have
\begin{align*}
t'(p)
&= t(p) + \dt\\
&\le t(u) + \dt\\
&\le t(u) + \e \minS \minW\\
&\le t(u) + \e \S(\fp'\fq\fr) \abs{up}
\end{align*}
because $\abs{up} \ge \minW$ and $\S(\fp'\fq\fr) \ge \minS$.  Since $0
< \e \le \half$ we have $\e \le \sqrt{1 - (1-\e)^2}$.
Therefore,
\begin{align*}
  t'(p)
&\le 
  t(u) + \abs{up} \sqrt{1 - (1-\e)^2} \S(\fp'\fq\fr)\\
&=
  t(u) + \abs{up} \sqrt{\S(\fp'\fq\fr)^2 
                        - (1-\e)^2 \S^2(\fp'\fq\fr)}\\
&\le
  t(u) + \abs{up} \sqrt{\S(\fp'\fq\fr)^2
                          - \norm{\grad \rest{t}{qr}}^2}
\end{align*}
which is precisely the causality constraint of
Equation~\ref{eqn:2d:causalityconstraint}.  The last inequality
follows because $\fp\fq\fr$ is progressive, hence $\norm{\grad
  \rest{t}{qr}} \le (1-\e) \S(\fp'\fq\fr) \phi_p \le (1-\e)
\S(\fp'\fq\fr)$.

% ..................................................................
\noindent\textbf{Case 2: $\mathbf{t(u) < t(q)}$}
See Figure~\ref{fig:pqr-Qobtuse}(a).  Let $\beta = \abs{uq}/\abs{up}$.
Since $\abs{uq} \ne 0$, we have
\begin{align}
  \frac{t'(p) - t(u)}{\abs{up}}
&= 
  \frac{t'(p) - t(q)}{\abs{up}} 
+ 
  \frac{t(q) - t(u)}{\abs{uq}} \frac{\abs{uq}}{\abs{up}}\nonumber\\
&= 
  \frac{t'(p) - t(q)}{\abs{up}} 
+ 
  \beta \norm{\grad \rest{t}{qr}}
\label{eqn:2d:strengthen}
\end{align}
Using Equation~\ref{eqn:2d:strengthen}, the causality constraint
(Equation~\ref{eqn:2d:causalityconstraint}) can be rewritten as
\begin{align}
  \frac{t'(p) - t(q)}{\abs{up}}
&\le 
  \sqrt{\S(\fp'\fq\fr)^2 
    - \norm{\grad \rest{t}{qr}}^2}\nonumber\\
&\quad {}-{} \beta \norm{\grad \rest{t}{qr}}
\label{eqn:2d:equiv_causalityconstraint}
\end{align}
Since $t_i$ is progressive, we have $\norm{\grad \rest{t}{qr}} \le
(1-\e) \S(\fp'\fq\fr) \phi_p$.  Substituting
this upper bound on $\norm{\grad \rest{t}{qr}}$ into
Equation~\ref{eqn:2d:equiv_causalityconstraint}, we obtain the
following constraint:
\begin{align}
  \frac{t'(p) - t(q)}{\abs{up}}
&\le 
  \S(\fp'\fq\fr) \left( \sqrt{1 - (1-\e)^2 \phi_p^2} \right)\nonumber\\
&\quad {}-{} \S(\fp'\fq\fr) (1-\e) \beta \phi_p
\label{eqn:2d:strong_causalityconstraint}
\end{align}
which implies the causality constraint of
Equation~\ref{eqn:2d:equiv_causalityconstraint}.
Now,
\begin{align*}
\frac{t'(p) - t(q)}{\abs{up}}
&\le 
  \frac{t'(p) - t(p)}{\abs{up}}\\
&\le
  \frac{\e \minS \minW}{\abs{up}}\\
&\le
  \e \minS.
\end{align*}
Since $\minS \le \S(\fp'\fq\fr)$,
Equation~\ref{eqn:2d:strong_causalityconstraint} is satisfied if
\[
  \e \le \sqrt{1 - (1-\e)^2 \phi_p^2} - (1-\e) \beta \phi_p
\]
or equivalently
\[
  \left( \e + (1-\e) \beta \phi_p \right)^2 
+
  (1-\e)^2 \phi_p^2
\le
  1
\]
We have
\[
  \left( \e + (1-\e) \beta \phi_p \right)^2 
+
  (1-\e)^2 \phi_p^2
=
  1 + 2 \e (1-\e) \left( \beta \phi_p - 1 \right)
\]
We have $\phi_p = \sin \angle{pqr} = \abs{up}/\abs{pq} >
\abs{up}/\abs{pr} = \sin \angle{prq}$ and $\beta = \abs{uq}/\abs{up}$.
Since $\abs{uq} < \abs{pq}$, we have $\beta \phi_{qr} < 1$.
Therefore, Equation~\ref{eqn:2d:strong_causalityconstraint} is
satisfied.
\end{proof}
% -----------------------------------------------------------------
% -----------------------------------------------------------------

% -----------------------------------------------------------------
% -----------------------------------------------------------------
\begin{theorem}
  If a front $t$ is progressive, then for any local minimum vertex $p$
  and for every $\dt \in [0,\minT]$ the front $t' = \next(t,p,\dt)$ is
  progressive.
\label{thm:2d:isprogressive}
\end{theorem}
\begin{proof}
Since only the triangles of the front incident on~$\fp$ advance
along with~$p$, we can restrict our attention to an arbitrary
triangle~$pqr$ incident on~$p$.  Let~$t$
and~$t'$ denote~$\rest{t_i}{pqr}$
and~$\rest{t_{i+1}}{pqr}$ respectively.
Let $u$ be the orthogonal projection of $p$ onto line $qr$.
Let $\progS$ denote $\S(\fp'\fq'\fr)$ where $\fp'=(p,t(p)+\dt)$ and
$\fq'=(q,t(q)+\minT)$.

We separate the analysis into three cases depending on which, if any,
of the angles $\angle{pqr}$ and $\angle{prq}$ of $\triangle{pqr}$ is
obtuse.

% ...............................................................
\noindent\textbf{Case 1: Both $\mathbf{\angle{pqr}}$ and
  $\mathbf{\angle{prq}}$ are non-obtuse.}
See Figure~\ref{fig:pqr-QRacute}(b).  In this case, we have $t(u) \ge
t(q) \ge t(p)$ and $\vec{n}_{qr} \cdot \vec{n}_{rp} \le 0$.  Let
$\alpha = \sqrt{1 - (\vec{n}_{qr} \cdot \vec{n}_{rp})^2}$.  Hence,
$\sin \angle{qrp} = \alpha = \abs{up}/\abs{pr}$ and
$\vec{n}_{qr} \cdot \vec{n}_{rp} = - \sqrt{1 - \alpha^2}
= - \abs{ur}/\abs{pr}$.  Also, $\phi_q = \max\,\{\sin \angle{qrp}, \sin
\angle{qpr}\}$.

Therefore, the progress constraint of
Equation~\ref{eqn:2d:progressconstraint} can be rewritten as follows:
\begin{equation}
\begin{array}{rcl}
  \frac{t'(p) - t(u)}{\abs{up}}
&\le&
  (1-\e) \S(\fp'\fq'\fr) 
  \frac{\max\{\sin \angle{qrp}, \sin \angle{qpr}\}}{\sin \angle{qrp}}\\[2ex]
&&
  {}+{} \frac{\abs{ur}}{\abs{up}} \, \norm{\grad \rest{t}{qr}}
\end{array}
\label{eqn:2d:equiv1_progressconstraint}
\end{equation}
We have
\[
\frac{t'(p) - t(u)}{\abs{up}}
\le 
  \frac{t'(p) - t(p)}{\abs{up}}\\
\le
  \frac{\e \minS \minW}{\abs{up}}\\
\le
  \e \minS.
\]
Since $\e \le \half$, we have $\e \le 1-\e$; also, $\minS \le
\S(\fp'\fq'\fr)$; hence,
\begin{align*}
\e \minS
&\le
  (1-\e) \S(\fp'\fq'\fr)\\
&\le
  (1-\e) \S(\fp'\fq'\fr) 
  \frac{\max\{\sin \angle{qrp}, \sin \angle{qpr}\}}{\sin \angle{qrp}}\\
&\le
  (1-\e) \S(\fp'\fq'\fr) 
  \frac{\max\{\sin \angle{qrp}, \sin \angle{qpr}\}}{\sin \angle{qrp}}\\
&\quad {}+{} \frac{\abs{ur}}{\abs{up}} \, \norm{\grad \rest{t}{qr}}
\end{align*}
Therefore, the progress constraint of
Equation~\ref{eqn:2d:equiv1_progressconstraint} is satisfied.

% ...............................................................
\noindent\textbf{Case 2: $\mathbf{\angle{pqr}}$ is obtuse.}
See Figure~\ref{fig:pqr-Qobtuse}(a).  In this case, we have $t(u) <
t(q)$ and $\vec{n}_{qr} \cdot \vec{n}_{rp} \le 0$.  Let $\alpha =
\sqrt{1 - (\vec{n}_{qr} \cdot \vec{n}_{rp})^2}$.  Hence, $\sin
\angle{qrp} = \alpha = \abs{up}/\abs{pr}$ and $\vec{n}_{qr} \cdot
\vec{n}_{rp} = - \sqrt{1 - \alpha^2} = - \abs{ur}/\abs{pr}$.

Let $\beta = \abs{uq}/\abs{up}$.  Since $\abs{uq} \ne 0$, we have
\begin{align*}
  \frac{t'(p) - t(u)}{\abs{up}}
&= 
  \frac{t'(p) - t(q)}{\abs{up}} 
+ 
  \frac{t(q) - t(u)}{\abs{uq}} \frac{\abs{uq}}{\abs{up}}\\
&= 
  \frac{t'(p) - t(q)}{\abs{up}} 
+ 
  \beta \norm{\grad \rest{t}{qr}}
\end{align*}
Therefore, the progress constraint of
Equation~\ref{eqn:2d:progressconstraint} can be rewritten as follows:
\begin{equation}
\begin{array}{rcl}
  \frac{t'(p) - t(q)}{\abs{up}}
&\le&
  (1-\e) \S(\fp'\fq'\fr) 
  \frac{\max\{\sin \angle{qrp}, \sin \angle{qpr}\}}{\sin \angle{qrp}}\\[2ex]
&&
  {}+{} \left( \frac{\abs{ur}}{\abs{up}} - \beta \right)
        \, \norm{\grad \rest{t}{qr}}
\end{array}
\label{eqn:2d:equiv2_progressconstraint}
\end{equation}
We have
\[
\frac{t'(p) - t(q)}{\abs{up}}
\le 
  \frac{t'(p) - t(p)}{\abs{up}}\\
\le
  \frac{\e \minS \minW}{\abs{up}}\\
\le
  \e \minS.
\]
Since $\e \le \half$, we have $\e \le 1-\e$; also, $\minS \le
\S(\fp'\fq'\fr)$; hence,
\begin{align*}
\e \minS
&\le
  (1-\e) \S(\fp'\fq'\fr)\\
&\le
  (1-\e) \S(\fp'\fq'\fr) 
  \frac{\max\{\sin \angle{qrp}, \sin \angle{qpr}\}}{\sin \angle{qrp}}\\
&\le
  (1-\e) \S(\fp'\fq'\fr) 
  \frac{\max\{\sin \angle{qrp}, \sin \angle{qpr}\}}{\sin \angle{qrp}}\\
&\quad {}+{} \left( \frac{\abs{ur}}{\abs{up}} - \beta \right)
        \, \norm{\grad \rest{t}{qr}}
\end{align*}
Therefore, the progress constraint of
Equation~\ref{eqn:2d:equiv2_progressconstraint} is satisfied.  The
last inequality follows because $\beta = \abs{uq}/\abs{up} <
\abs{ur}/\abs{up}$.

% ...............................................................
\noindent\textbf{Case 3: $\mathbf{\angle{prq}}$ is obtuse.}
See Figure~\ref{fig:pqr-Robtuse}(c).  In this case, we have $t(u) \ge
t(r) \ge t(q) \ge t(p)$ and $\vec{n}_{qr} \cdot \vec{n}_{rp} > 0$.
Let $\alpha = \sqrt{1 - (\vec{n}_{qr} \cdot \vec{n}_{rp})^2}$.  Hence,
$\sin \angle{qrp} = \alpha = \abs{up}/\abs{pr}$ and $\vec{n}_{qr}
\cdot \vec{n}_{rp} = \sqrt{1 - \alpha^2} = \abs{ur}/\abs{pr}$.

Let $\beta = \abs{uq}/\abs{up}$.  Since $\abs{uq} \ne 0$, we have
\begin{align*}
  \frac{t'(p) - t(u)}{\abs{up}}
&= 
  \frac{t'(p) - t(q)}{\abs{up}} 
+ 
  \frac{t(q) - t(u)}{\abs{uq}} \frac{\abs{uq}}{\abs{up}}\\
&= 
  \frac{t'(p) - t(q)}{\abs{up}} 
- 
  \beta \norm{\grad \rest{t}{qr}}
\end{align*}
Therefore, the progress constraint of
Equation~\ref{eqn:2d:progressconstraint} can be rewritten as follows:
\begin{equation}
\begin{array}{rcl}
  \frac{t'(p) - t(q)}{\abs{up}}
&\le&
  (1-\e) \S(\fp'\fq'\fr) 
  \frac{\max\{\sin \angle{qrp}, \sin \angle{qpr}\}}{\sin \angle{qrp}}\\[2ex]
&&
  {}+{} \left( \beta - \frac{\abs{ur}}{\abs{up}} \right)
        \, \norm{\grad \rest{t}{qr}}
\end{array}
\label{eqn:2d:equiv3_progressconstraint}
\end{equation}
As before, we have
\[
\frac{t'(p) - t(q)}{\abs{up}}
\le 
  \frac{t'(p) - t(p)}{\abs{up}}\\
\le
  \frac{\e \minS \minW}{\abs{up}}\\
\le
  \e \minS.
\]
Since $\e \le \half$, we have $\e \le 1-\e$; also, $\minS \le
\S(\fp'\fq'\fr)$; hence,
\begin{align*}
\e \minS
&\le
  (1-\e) \S(\fp'\fq'\fr)\\
&\le
  (1-\e) \S(\fp'\fq'\fr) 
  \frac{\max\{\sin \angle{qrp}, \sin \angle{qpr}\}}{\sin \angle{qrp}}\\
&\le
  (1-\e) \S(\fp'\fq'\fr) 
  \frac{\max\{\sin \angle{qrp}, \sin \angle{qpr}\}}{\sin \angle{qrp}}\\
&\quad 
  {}+{} \left( \beta - \frac{\abs{ur}}{\abs{up}} \right)
        \, \norm{\grad \rest{t}{qr}}
\end{align*}
Therefore, the progress constraint of
Equation~\ref{eqn:2d:equiv3_progressconstraint} is satisfied.  The
last inequality follows because $\beta = \abs{uq}/\abs{up} >
\abs{ur}/\abs{up}$.
\end{proof}
% -----------------------------------------------------------------
% -----------------------------------------------------------------

% ................................................................
\begin{theorem}
  For any $i \ge 0$, if the front $t_i$ is progressive then $t_i$ is
  valid.
\label{thm:2d:progressiveisvalid}
\end{theorem}

The proof is almost identical to that of Theorem~\ref{thm:1d:causalisvalid}.

% ................................................................

% ================================================================
% ================================================================
% ================================================================
\subsection{Being greedy}

We would like to maximize the progress at each step in a greedy
fashion, i.e., given a front $t_i$ we would like to maximize
$t_{i+1}(p)$ where $t_{i+1} = \next(t_i,p,\dt)$ subject to the
constraint that $t_{i+1}$ is causal.  For a fixed triangle
$pp_1p_2{\ldots}p_d$ incident on $p$ let $\supT^{i+1}$ denote
$\sup\,\{T$ : $\fp'\fq\fr$ is causal and progressive, where
$\fp'=(p,T)$ and $\fp_1'=(p_1,t_i(p_1)+\minT\}$.  To maximizing the
progress at step $i+1$, we would like to compute $\supT^{i+1}$.
Similar to the 1D$\times$Time case, partition the set of cones of
influence from points on the front $t_i$ into local and remote
subsets.  Let $\S_{\text{local}}$ denote the smallest slope among all
local cones of influence.  The triangle $\fp'\fq\fr$ is causal only if
its slope is less than or equal to $\S_{\text{local}}$.  Let
$T_{\text{local}}$ be the maximum time value of $\fp'$ for which the
slope of $\fp'\fq\fr$ is less than or equal to $\S_{\text{local}}$.
The maximum $T_{\text{local}}$ exists because the set of allowed
values of $T$ where $\fp'=(p,T)$ is closed and therefore compact.  To
compute $T_{\text{local}}$ we substitute $\S_{\text{local}}$ in the
condition for causality of $\fp'\fq\fr$.

Unlike the 1D$\times$Time case, it is not clear that $\supT^{i+1}$ can
be computed by ray shooting queries.  In 2D$\times$Time, we need an
oracle to determine which among several right circular cones is
intersected first by a triangle $\fp'\fq'\fr$ when the vertex $\fp$ of
$\triangle{\fp\fq\fr}$ is lifted to $\fp'=(p,T)$ while also lifting
$\fq$ to $\fq'=(q,t(q)+\minT)$.  However, just as for the
1D$\times$Time case, we can approximate $\supT^{i+1}$ up to any given
numerical accuracy by performing a binary search in the interval
$[t_i(p)+\minT, T_{\text{local}}]$ which we know contains
$\supT^{i+1}$.  Therefore, the eventual height of the tentpole
$\fp\fp'$ is at least $\max\{\minT, \supT^{i+1} - \eta\}$ where $\eta
> 0$ is the desired numerical accuracy.

%\bigskip

We thus have the following theorem.

% ..................................................................
\begin{theorem}
  Given a triangulation $\sp$ of a bounded planar space domain
  where $\minW$ is the minimum width of a simplex of $\sp$ and $\minS$
  is the minimum slope anywhere in $\sp \times [0,\infty)$, for
  every~$\e$ such that~$0 < \e \le \half$ our algorithm constructs a
  simplicial mesh of~$\sp \times [0,T]$ consisting of at
  most~$\ceil{\frac{\diam(\sp) \, \maxS \, \Delta(\sp)}{\e \minS
      \minW} \, T}$ spacetime elements for every real $T \ge 0$.
\label{thm:2d:main}
\end{theorem}
\begin{proof}
  By Theorems~\ref{thm:2d:iscausal} and~\ref{thm:2d:isprogressive}, it
  follows that the height of each tentpole constructed by the
  algorithm is at least $\minT = \e \minS \minW$.  By
  Theorem~\ref{thm:2d:progressiveisvalid}, after constructing at most
  $k \le \ceil{\frac{\diam(\sp) \, \maxS}{\minT} \, T}$ patches, the
  entire front $t_k$ is past the target time $T$.  Since each patch
  consists of at most $\Delta(\sp)$ elements, the theorem follows.
\end{proof}
% ..................................................................

%% End algorithm for 2D x Time.

% #################################################################
% #################################################################
% #################################################################

\section{Conclusion}
\label{sec:conclusion}

We have shown how to extend the Tent Pitcher algorithm for planar and
linear spatial domains to the case of changing wavespeeds.  Our
expressions for the causality and progress constraints that apply at
each step make explicit the dependence on the slope of the cone of
influence most constraining the progress at that step.  This
dependence is not explicit in the formul\ae{} of Erickson \etal{}
because they assume without loss of generality that the slope is~$1$
everywhere in spacetime.  For the constant wavespeed case, the
algorithm in this paper is an alternative to the algorithm due to
Erickson \etal{} with potentially weaker progress constraints.  We can
view the algorithm of Erickson \etal{} as looking one step ahead in
the sense that the progress constraint at step $i$ guarantees that the
front constructed in step $i+1$ is causal.  Our algorithm can be
viewed as looking one step even further---our progress constraint at
step $i$ guarantees that the front constructed in step $i+2$ is
causal.  In a relatively straightforward manner, we can generalize
this idea to looking at step $i$ to the front in step $i+h$ where $h$
is a \emph{horizon} parameter that can be chosen adaptively by the
algorithm.  It needs to be investigated whether the extra complexity
of the algorithm for $h > 2$ is justified by a more efficient meshing
algorithm overall.

We have preliminary experimental results in 1D$\times$Time and a
prototype with simulated physics in 2D$\times$Time; more substantial
empirical study is required and we expect to report results of such a
study soon.  One of the objectives of the study will be to explore
different heuristics to choose which local minimum vertex to pitch at
every step.  Some heuristics, such as pitching the local minimum with
the minimum slope (highest wavespeed), perform better than others.  We
have an extension to the current algorithm that allows pitching at any
vertex, not necessarily a local minimum.  However, the extended
algorithm is more complicated and it is not clear if the expected
gains will be worth the extra computation time.

\begin{figure}\centering\sf
\iffig\includegraphics[width=0.45\textwidth]{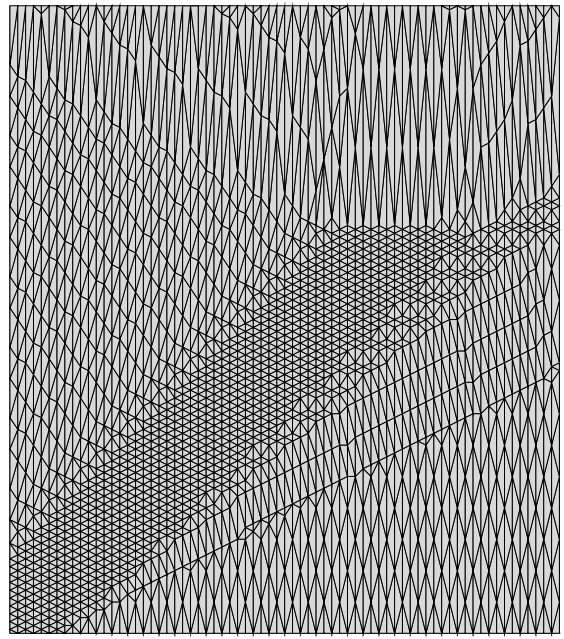}\fi
\caption{An unstructured triangular spacetime mesh over a 1D uniform
space mesh.  The space dimension is horizontal and time increases
upwards.  The slope at any point in spacetime is one of three distinct
values: the minimum slope occurs in a band around the diagonal where
the tentpoles are shortest; beyond a certain time value, the maximum
slope occurs everywhere.}
\label{fig:1d:example}
\end{figure}

\begin{figure}\centering\sf
\iffig\includegraphics[angle=90,width=0.48\textwidth]{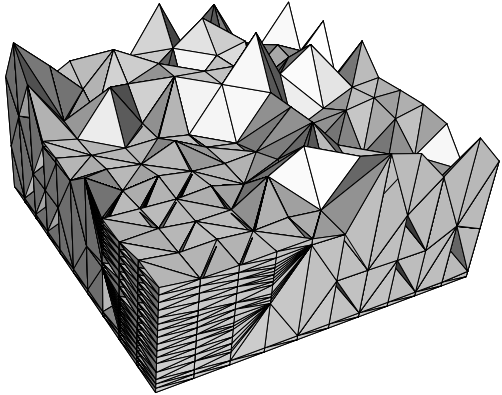}\fi
\caption{An unstructured tetrahedral spacetime mesh over a triangulated
  uniform 2D grid.
  Time increases upwards.
  The slope at any point in spacetime is one of two distinct
  values: the minimum slope occurs inside a circular cone where the
  tentpoles are shortest, the maximum slope occurs everywhere else.}
\label{fig:2d:example}
\end{figure}

Figures~\ref{fig:1d:example} and~\ref{fig:2d:example} illustrate
spacetime meshes constructed by our prototype implementation over 1D
and 2D space meshes respectively.  The 1D$\times$Time spacetime mesh
was constructed by pitching an independent set of local minima in
non-increasing order of wavespeed.  In other words, the algorithm
preferred to pitch local minima adjacent to points on the front where
the wavespeed was maximum (slope was minimum).  The 2D$\times$Time
mesh was constructed by pitching a global minimum at every step.  In
either example, many more spacetime elements would be required to mesh
the same volume if the height of every tentpole were constrained by
the globally minimum slope.

% ................................................................

% ................................................................

In higher dimensions, we have a theorem identical to
Theorem~\ref{thm:2d:main} when every dihedral angle of every simplex
is non-obtuse.  We anticipate soon an analogous theorem for arbitrary
dimensional space domains in the presence of obtuse angles.

Our algorithm can be modified to handle asymmetric cones, such as due
to wave propagation through anisotropic media.  In the presence of
anisotropy, the most limiting cone constraint can be nonlocal.

In a recent paper, Abedi \etal{}~\cite{abedi04spacetime} extend
TentPitcher to support another kind of adaptivity, where the size of
the spacetime elements is adapted to \aposteriori{} estimates of the
numerical error.  Abedi \etal{} apply hierarchical refinement and
coarsening of the underlying one- or two-dimensional space mesh to
adapt the spatial size of future spacetime elements.  They extend the
progress constraints of Erickson \etal{} to anticipate future
refinement and coarsening both of which change the shape of the
elements on the front.  The outstanding problem that we plan to
consider next is to combine adaptivity to changing wavespeeds with
refinement and coarsening for the case of planar space domains.  It is
quite straightforward to combine the progress constraints in this
paper with those of Abedi \etal{} to support refinement in the
presence of changing wavespeeds.  Coarsening can be done safely if
each triangle after coarsening satisfies progress constraint
[$\minS$].  When coarsening is possible only under such strict
constraints, we need to carefully prioritize each coarsening step so
that the front is only as refined as necessary and not much more.

Our research group is also implementing a parallel version of Tent
Pitcher to run on multiple processors.  The nonlocal nature of the
constraints pose significant challenges in the parallel setting.

In many problems, the geometry of the space domain changes over time.
There may also be internal boundaries between different parts of the
domain, e.g., separating two distinct materials with different
physical properties, and these internal boundaries may evolve over
time.  We would like to handle moving boundaries both internal and
external.

%% End conclusion.

% #################################################################
\paragraph{Acknowledgments}

The author would like to thank the other members of the CPSD spacetime
meshing group, especially Jeff Erickson, Yong Fan, Robert Haber, Mark
Hills, and Jayandran Palaniappan.  Thanks also to the anonymous
referee, whose comments were especially useful.

% #################################################################
% #################################################################
% #################################################################

\bibliography{spacetime}

\begin{thebibliography}{10}
\newcommand{\enquote}[1]{``#1''}
\expandafter\ifx\csname url\endcsname\relax
  \def\url#1{\texttt{#1}}\fi
\expandafter\ifx\csname urlprefix\endcsname\relax\def\urlprefix{URL }\fi

\bibitem{abedi04spacetime}
Abedi R., Chung S.H., Erickson J., Fan Y., Garland M., Guoy D., Haber R.,
  Sullivan J.M., Thite S., Zhou Y.
\newblock \enquote{Spacetime Meshing with Adaptive Refinement and Coarsening.}
\newblock \emph{Proc. 20th Symp. Computational Geometry}, pp. 300--309. June
  2004

\bibitem{barnes-hut86nbody}
Barnes J.E., Hut P.
\newblock \enquote{A Hierarchical $O(N Log N)$ Force Calculation Algorithm.}
\newblock \emph{Nature}, vol. 324, no.~4, 446--449, December 1986

\bibitem{CockburnKS00}
Cockburn B., Karniadakis G., Shu C.
\newblock \emph{Discontinuous {Galerkin} methods: theory, computation and
  applications}, vol.~11 of \emph{Lecture Notes in Computational Science and
  Engineering}.
\newblock Springer, 2000

\bibitem{erickson02building}
Erickson J., Guoy D., Sullivan J.M., {\"U}ng{\"o}r A.
\newblock \enquote{Building Space-Time Meshes over Arbitrary Spatial Domains.}
\newblock \emph{Proc. 11th Int'l. Meshing Roundtable}, pp. 391--402. 2002

\bibitem{guttman84rtrees}
Guttman A.
\newblock \enquote{A Dynamic Index Structure for Spatial Searching.}
\newblock \emph{Proc. ACM SIGMOD Conf. Principles Database Systems}, pp.
  47--57. 1984

\bibitem{lin96collision}
Lin M., Manocha D., Cohen J., Gottschalk S.
\newblock \emph{Algorithms for Robotics Motion and Manipulation}, chap.
  Collision Detection: Algorithms and Applications, pp. 129--142.
\newblock A.K. Peters, 1996

\bibitem{Lowrie98}
Lowrie R.B., Roe P.L., van Leer B.
\newblock \enquote{Space-Time Methods for Hyperbolic Conservation Laws.}
\newblock \emph{Barriers and Challenges in Computational Fluid Dynamics},
  vol.~6 of \emph{{ICASE}/{LaRC} Interdisciplinary Series in Science and
  Engineering}, pp. 79--98. Kluwer, 1998

\bibitem{Richter94}
Richter G.R.
\newblock \enquote{An explicit finite element method for the wave equation.}
\newblock \emph{Applied Numerical Mathematics}, vol.~16, 65--80, 1994

\bibitem{ungor02phd}
\"Ung\"or A.
\newblock \emph{Parallel {D}elaunay Refinement and Space-Time Meshing}.
\newblock Ph.D. thesis, University of Illinois at Urbana-Champaign, October
  2002

\bibitem{ungor00tentpitcher}
{\"U}ng{\"o}r A., Sheffer A.
\newblock \enquote{{T}ent-{P}itcher: A Meshing Algorithm for Space-Time
  Discontinuous {Galerkin} Methods.}
\newblock \emph{Proc. 9th Int'l. Meshing Roundtable}, pp. 111--122. 2000

\bibitem{YinASHT00}
Yin L., Acharya A., Sobh N., Haber R., Tortorelli D.A.
\newblock \enquote{A Space-time discontinuous {Galerkin} method for
  elastodynamic analysis.}
\newblock B.~Cockburn, G.~Karniadakis, C.~Shu, editors, \emph{Lecture Notes in
  Computational Science and Engineering}, vol.~11, pp. 459--464. Springer, 2000

\end{thebibliography}

%---
\end{document}